\documentstyle[epsfig]{elsart}
\begin{document}

\newcommand{\re}{\mathop{\mathrm{Re}}}
\newcommand{\im}{\mathop{\mathrm{Im}}}
\newcommand{\D}{\mathop{\mathrm{d}}}
\newcommand{\I}{\mathop{\mathrm{i}}}
\newcommand{\E}{\mathop{\mathrm{e}}}

\noindent {\Large DESY 04-013}

\noindent {\Large January 2004}

\bigskip

\begin{frontmatter}

\journal{Optics Communications}

\title{
Terawatt-scale sub-10-fs laser technology --
key to generation of GW-level
attosecond pulses in X-ray free electron laser }

\author{E.L.~Saldin},
\author{E.A.~Schneidmiller},
and
\author{M.V.~Yurkov}

\address{Deutsches Elektronen-Synchrotron (DESY),
Hamburg, Germany}

\begin{abstract}

We propose a technique for the production of attosecond X-ray pulses
which is based on the use of X-ray SASE FEL combined with a femtosecond
laser system. A few-cycle optical pulse from a Ti:sapphire laser
interacts with the electron beam in a two-period undulator resonant
to 800 nm wavelength and produces energy modulation within a slice of
the electron bunch. Following the energy modulator the electron beam
enters the X-ray undulator and produces SASE radiation. Due to energy
modulation the frequency is correlated to the longitudinal position
within the few-cycle-driven slice of SASE radiation pulse. The largest
frequency offset corresponds to a single-spike pulse in the time domain
which is confined to one half-oscillation period near the central peak
electron energy. The selection of single-spike pulses is achieved by
using a crystal monochromator after the X-ray undulator. Our studies
show that the proposed technique is capable to produce 300 attoseconds
long single pulses with GW-level output power in the 0.1~nm wavelength
range, and is applicable to the European X-Ray Laser Project XFEL and
the Linac Coherent Light Source at SLAC.

\end{abstract}

\end{frontmatter}

\clearpage

$\mbox{}$

\thispagestyle{empty}

\clearpage

\setcounter{page}{1}

\section{Introduction}

At the start of this century, we have seen a revolution in synchrotron
radiation source intensities. This revolution stemmed from the
technique of free electron laser (FEL) based on self-amplified
spontaneous emission (SASE), combined with recent progress in
accelerator technologies, developed in connection with high-energy
linear colliders. In 2001, the VUV FEL at the TESLA Test Facility at
DESY (Hamburg, Germany) has successfully demonstrated saturation from
82 nm to 125 nm with GW-level peak power and pulse duration down to
40~fs \cite{ay1,ay2}. It is the first result from this device that
Wabnitz et al. reported in \cite{w}. They illuminated xenon clusters
with high-intensity ($10^{14}$~W/cm$^{2}$) VUV FEL pulses and observed
an unexpectedly strong absorption of the VUV radiation. Such a highly
nonlinear optical interaction between light and matter at VUV
wavelength range has never been seen before and these fascinating
results show the potential of this new class of light sources for
scientific research.  While modern third generation synchrotron light
sources are reaching their fundamental performance limit, recent
successes in the development of the VUV FEL at DESY have paved the way
for the construction of the novel type of light source which will
combine most of the positive aspects of both a laser and a synchrotron.
Starting in 2004, the phase 2 extension of TTF will deliver FEL
radiation down to the soft X-ray spectral range with minimum wavelength
of about 6 nm in the first harmonic and reaching into "water window" in
the second harmonic \cite{ttf2}.

Recently the German government, encouraged by these results, approved
funding a hard X-ray SASE FEL user facility -- the European X-Ray Laser
Project XFEL. The US Department of Energy (DOE) has given SLAC the
goahead for engineering design of the  Linac Coherent Light Source
(LCLS) device to be constructed at SLAC.  These devices should produce
100 fs X-ray pulses with over 10 GW of peak power \cite{tdr1,tdr2}.
These new X-ray sources will be able to produce intensities of the
order of $10^{18} \ \mathrm{W/cm}^{2}$. The main difference between the
two projects is the linear accelerator, an existing room temperature
linac for LCLS at SLAC, and a future superconducting linac for European
XFEL. The XFEL based on superconducting accelerator technology will
make possible not only a jump in a peak brilliance by ten orders of
magnitude, but also an increase in average brilliance by five orders of
magnitude compared to modern 3rd generation synchrotron radiation
facilities. The LCLS and European XFEL projects are scheduled to start
operation in 2008 and 2012, respectively.

The motivation for the development of XFELs was recently described in
detail in \cite{tdr1,tdr2}.  The discussion in the scientific community
over the past decade has produced many ideas for novel applications of
the X-ray laser.  Brilliance, coherence, and timing down to the
femtosecond regime are the three properties which have the highest
potential for new science to be explored with an XFEL. In its initial
configuration the XFEL pulse duration is about 100 femtoseconds. Even
though this is a few hundreds times shorter than in third generation
light sources, it can probably be further reduced to about 10
femtoseconds \cite{fs1,sb,fs2}. A novel way to generate sub-10 fs x-ray
pulses -- the slotted spoiler method (P. Emma, 2003) has been proposed
recently. This method is based on spoiling the beam phase density in a
part of the electron bunch so that this part will not lase, while
preserving lasing in a short length of the bunch. The FEL performance
of the spoiled beam approach was computed using the time-dependent
GENESIS simulation. It has been shown that it is possible to produce
X-ray pulses with duration of 3-4 fs FWHM for nominal LCLS bunch
compression parameters \cite{fs3}.

Femtosecond-resolution experiments with X-rays can possibly show
directly how matter is formed out of atoms. In fact, X-ray pulse
duration even shorter than one femtosecond may be useful for many
scientific applications. The reason is that phenomena inside atoms
occur on sub-femtosecond timescale. Generating single attosecond $\sim
0.1$~nm X-ray pulses is one of the biggest challenges in physics. The
use of such a tool will enable to trace process inside the atoms for
the first time. If there is any place where we have a chance to test
the main principles of quantum mechanics in the pure way, it is there.

The interest in the science with attosecond pulses is growing rapidly
in the very large laser community. This community is familiar with
attosecond pulses of light at photon energies up to about 0.5~keV
(3~nm). This is achieved by focusing a fs laser into a gas target
creating radiation of high harmonics of fundamental laser frequency.
The key to these developments was the invention of laser systems
delivering pulses in the range of 5~fs with pulse energies higher than
a fraction of mJ. This approach produced the first measurable XUV
pulses in the 200~as regime \cite{laser-atto1,laser-atto2}. In
principle, table-top ultra-fast X-ray sources have the right duration
to provide us with a view of subatomic transformation processes.
However, their power and photon energy are by far low. The XFEL is
ideally suited the purpose of this emerging field of science. Recently
an approach for the generation of attosecond pulses combining fs
quantum laser and harmonic cascade (HC) FEL scheme
\cite{yu-hghg,bill-hc} was proposed in \cite{bill-sasha}. The HC FEL
scheme has the potential to produce coherent light down to wavelengths
of a few nm in an undulator sequence \cite{limit-hghg}. The analysis
presented in \cite{bill-sasha} shows that this technique has potential
to produce 100~as long radiation pulses with MW-level of output power
down to 1~nm wavelength.

The X-ray SASE FEL holds a great promise as a source of radiation for
generating high power, single attosecond pulses. What ultimately limits
the XFEL pulse duration? Since the temporal and spectral
characteristics of the radiation field are related to each other
through Fourier transform, the bandwidth of the XFEL and the pulse
duration cannot vary independently of each other. There is a minimum
duration-bandwidth product (uncertainty principle). The shortest
possible X-ray pulse duration generated by XFEL is limited by the
intrinsic bandwidth of the SASE process.  In the case of the European
XFEL and the LCLS, the FWHM bandwidth near saturation (at 0.1 nm) is
about 0.1\%, indicating a 300-as coherence time determined by the
bandwidth product. Recently a scheme to achieve pulse durations down to
400-600 attoseconds at a wavelength of 0.1 nm has been proposed
\cite{as}. It uses a statistical properties of SASE FEL high harmonic
radiation. The selection of a single 10-GW level attosecond pulses is
achieved by using a special trigger in data acquisition system. A
promising scheme for attophysics experiments using this approach has
been studied and could be implemented in the XFEL design \cite{sapp}.

In this paper we propose a new method allowing to reduce the pulse
length of the X-ray SASE FEL to the shortest conceptual limit of about
300~as. It is based on the application of a sub-10-fs laser for slice
energy modulation of the electron beam, and application of a crystal
monochromator for the selection of single attosecond pulses with
GW-level output power.

\section{The principle of attosecond techniques based on the use of
XFEL combined with fs quantum laser}

\begin{figure}[b]

\vspace*{0mm}

\begin{center}
\epsfig{file=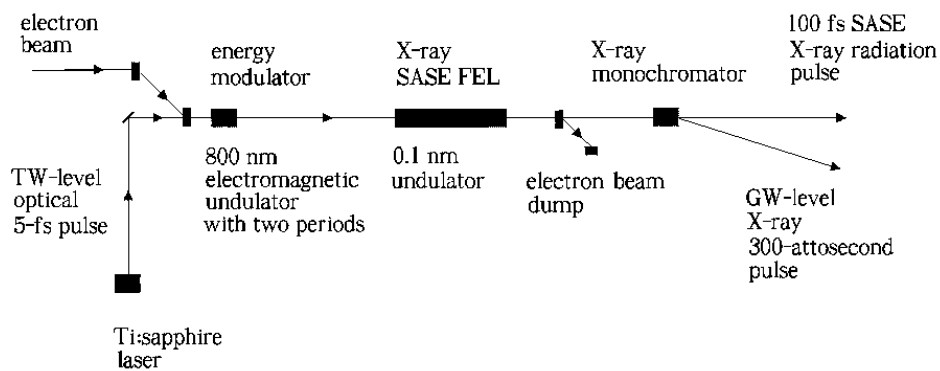,width=0.9\textwidth}
\end{center}
\caption{Schematic diagram of attosecond X-ray source}
\label{fig:xas1}
\end{figure}

\begin{figure}[tb]
\begin{center}
\epsfig{file=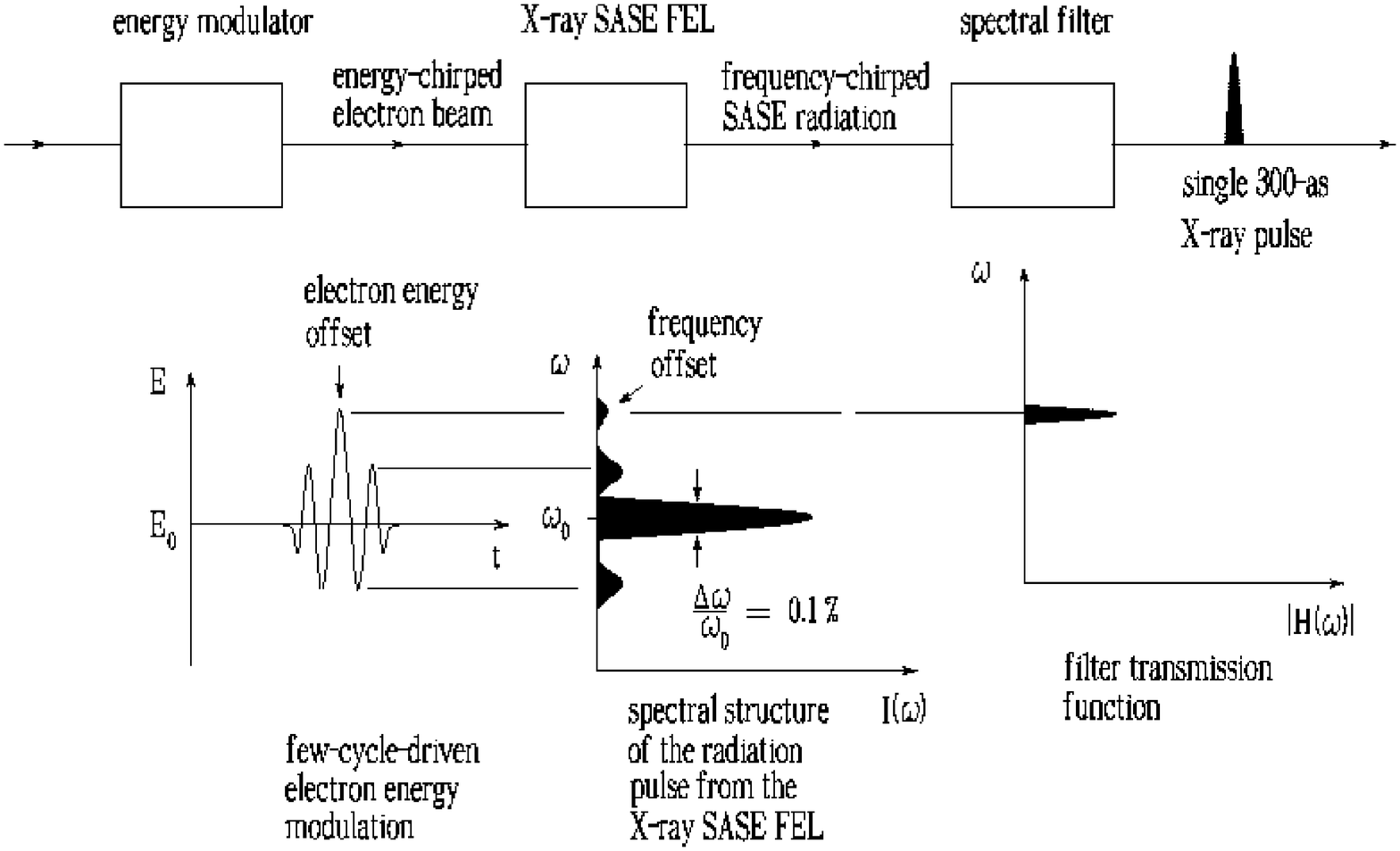,width=0.75\textwidth}
\end{center}
\caption{Sketch of single attosecond X-ray pulse synthesation through
frequency chirping and spectral filtering}
\label{fig:xas2}

\begin{center}
\epsfig{file=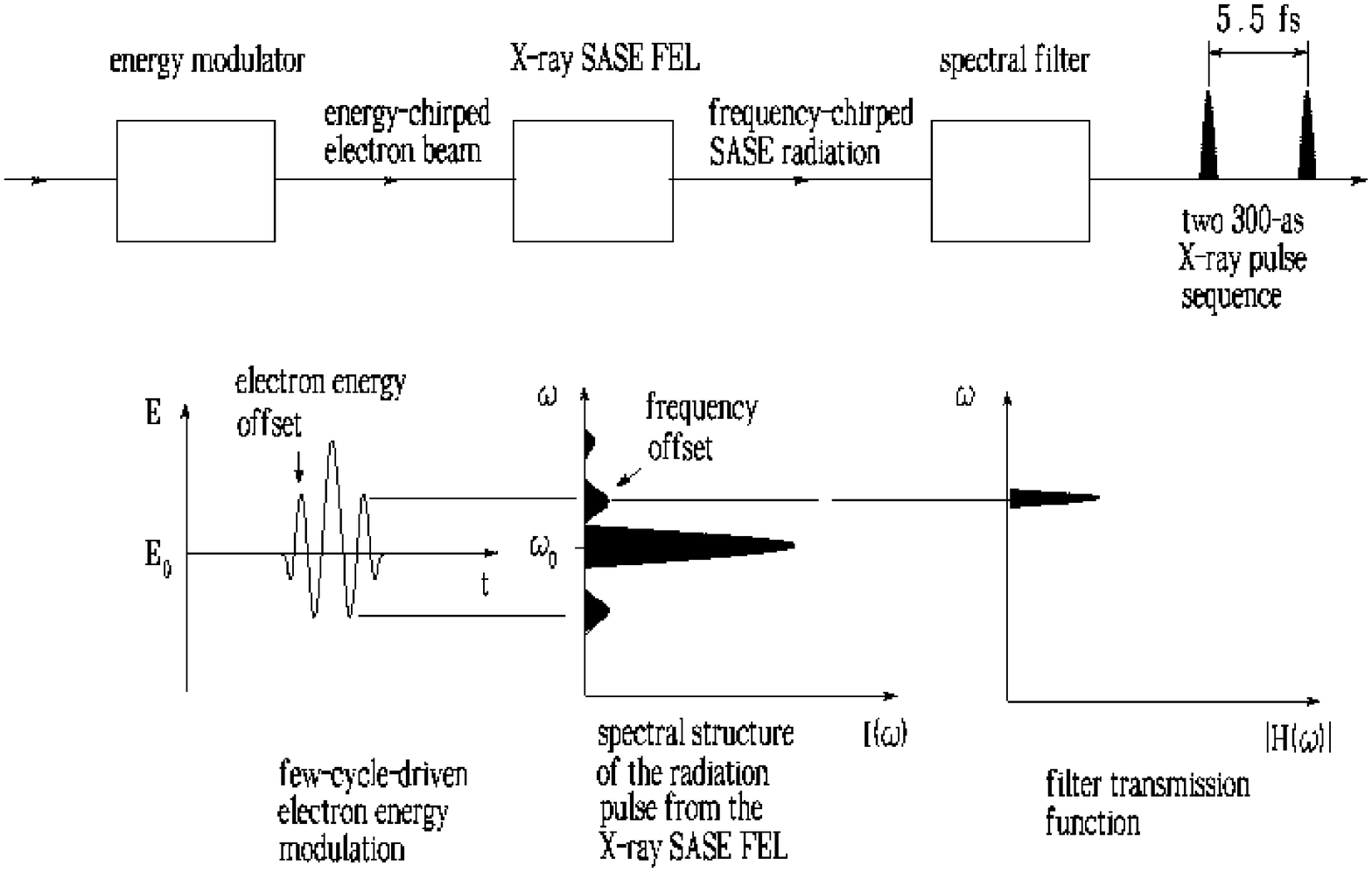,width=0.75\textwidth}
\end{center}
\caption{Sketch of two attosecond X-ray pulse sequence synthesation
through frequency chirping and spectral filtering. Pulse separation is
$2T_{0}$, where $T_{0}$ is the Ti:sapphire laser oscillation period}
\label{fig:xas3}
\end{figure}

A basic scheme of the attosecond X-ray source is shown in
Fig.~\ref{fig:xas1}. An ultrashort laser pulse is used to modulate the
energy of electrons within the femtosecond slice of the electron bunch
at the seed laser frequency. The seed laser pulse will be timed to
overlap with the central area of the electron bunch. It serves as a
seed for a modulator which consists of a short (a few periods)
undulator. Following the energy modulator the beam enters the X-ray
undulator. The process of amplification of radiation in this undulator
develops in the same way as in a conventional X-ray SASE FEL:
fluctuations of the electron beam current serve as the input signal
\cite{book}. The proposed scheme for the generation of attosecond
pulses is based on frequency-chirping the SASE radiation pulse. When an
electron beam traverses an undulator, it emits radiation at the
resonance wavelength $\lambda =
\lambda_{\mathrm{w}}(1+K^{2}/2)/(2\gamma^{2})$. Here
$\lambda_{\mathrm{w}}$ is the undulator period, $mc^{2}\gamma$ is the
electron beam energy, and $K$ is the undulator parameter.  The
laser-driven sinusoidal energy chirp produces a correlated frequency
chirp of the resonant radiation $\delta\omega/\omega \simeq
2\delta\gamma/\gamma$. After the undulator, the radiation is passed
through a crystal monochromator which reflects a narrow bandwidth.
Since the radiation frequency is correlated to the longitudinal
position within the beam, a short temporal radiation pulse is
transmitted through the monochromator.

Recent technological advances in ultrafast optics have permitted the
generation of optical pulses comprising only a few oscillation cycles
of the electric and magnetic fields. The pulses are delivered in a
diffraction-limited beam \cite{rmp}. The combination of a X-ray SASE
FEL and a few-cycle laser field techniques is very promising.  Our
concept of an attosecond X-ray facility is based on the use of a
few-cycle optical pulse from a Ti:sapphire laser system.  This optical
pulse is used for the modulation of the energy of the electrons within
a slice of the electron bunch at a wavelength of 800 nm.  Due to the
extreme temporal confinement, moderate optical pulse energies of the
order of a few mJ can result in an electron energy modulation amplitude
larger than 30-40 MeV. In few-cycle laser fields high intensities can
be "switched on" nonadiabatically within a few optical periods. As a
result, a central peak electron energy modulation is larger than other
peaks. This relative energy difference is used for the selection of
SASE radiation pulses with a single spike in the time domain by means
of a crystal monochromator. A schematic, illustrating these processes,
is shown in Fig. \ref{fig:xas2}.  Many different output fields can be
realized by using different spectral windowing. For instance, it is
possible to generate a sequence of 300-as X-ray pulses, separated by
$T_{0}$ (or $2T_{0}$), where $T_{0}$ is the Ti:sapphire laser
oscillation period. Such operation of the attosecond X-ray source is
illustrated in Fig. \ref{fig:xas3}.

The discussion in this paper is focused on the parameters for the
European XFEL operating in the wavelength range around 0.1~nm
\cite{tdr1}. Optimization of the attosecond SASE FEL has been performed
with the three-dimensional, time dependent code FAST \cite{fast} taking
into account all physical effects influencing the SASE FEL operation
(diffraction effects, energy spread, emittance, slippage effect, etc.).
In our scheme the separation of the frequency offset from the central
frequency by a monochromator is used to distinguish the 300-as pulse
from the 100 fs intense SASE pulse. The monochromatization is
straightforward: for the 0.1~nm wavelength range, Bragg diffraction is
the main tool used for such purposes.  In this case, one has to take
care that the short pulse duration is preserved. Transmission through
the monochromator will produce some intrinsic spreading of the pulse,
and the minimum pulse duration which may be selected by this method is
limited by the uncertainty principle.  The number of possible
reflections which provide the required spectral width is rather
limited. We are discussing here only Ge crystals, which have the
largest relative bandwidth.  This is an important feature which ensures
the preservation of the single-spike pulse duration. In its simplest
configuration the monochromator consists of Ge crystal diffracting from
the (111) lattice planes. We show that it is possible to produce X-ray
pulses with FWHM duration of 300 as. In some experimental situations
this simplest configuration of monochromator is not optimal. In
particular, our study has shown that the maximum contrast of the
attosecond X-ray pulses does not exceed 80\% and is due to the long
tail of the intrinsic crystal reflectivity curve.  The obvious and
technically possible solution of the problem of contrast increase might
be to use a premonochromator. One can align the premonochromator so
that the main peak of the spectrum is blocked.

\section{Generation of attosecond pulses from XFEL}

In the following we illustrate the operation of an attosecond SASE FEL
for the parameters close to those of the European XFEL operating at the
wavelength 0.1 nm \cite{tdr1}. The parameters of the electron beam are:
energy 15~GeV, charge 1~nC, rms pulse length 25~$\mu $m, rms normalized
emittance 1.4~mm-mrad, rms energy spread 1~MeV. Undulator period is
3.4~cm.

\subsection{Slice modulation of the electron beam}

\begin{figure}[b]
\begin{center}

\vspace*{-8mm}

\epsfig{file=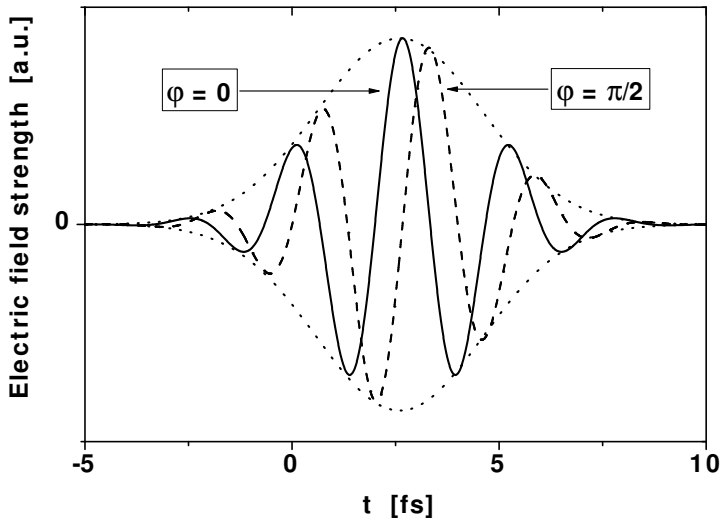,width=0.8\textwidth}

\vspace*{-8mm}

\end{center}
\caption{
Possible evolutions of the electric field in the 5-fs pulse.
carried at a wavelength 800 nm for two different pulse phases ($\phi =
0, \pi/2$)
}
\label{fig:lf}
\end{figure}

The parameters of the seed laser are: wavelength 800 nm, energy in
the laser pulse 2--4 mJ, and FWHM pulse duration 5 fs (see Fig.
\ref{fig:lf}). The laser beam is focused onto the electron beam in a
short undulator resonant at the
optical wavelength of 800~nm. Parameters of the undulator are:
period length 50~cm, peak field 1.6~T, number of periods 2.

\begin{figure}[tb]
\begin{center}

\vspace*{-8mm}

\epsfig{file=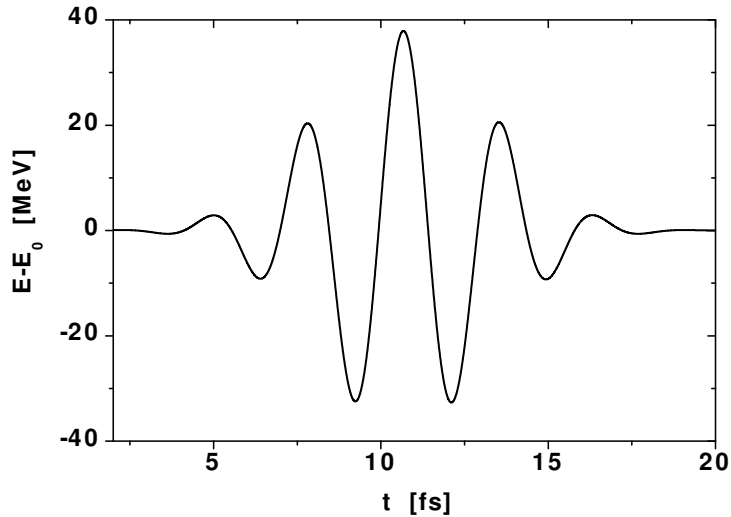,width=0.8\textwidth}

\vspace*{-8mm}

\end{center}
\caption{
Energy modulation of the electron beam at the exit of the
modulator undulator. The laser parameters are $\lambda =$ 800 nm,
$W_{\mathrm{peak}} =$ 800 GW, and  FWHM pulse duration of
$\tau_{\mathrm{p}} =$ 5 fs
}
\label{fig:emod40}
\end{figure}

Optimal conditions of the focusing correspond to the positioning of the
laser beam waist in the center of the undulator. In laser pulses
comprising just a few wave cycles, the amplitude envelope and the
carrier frequency are not sufficient to characterize and control laser
radiation, because the evolution of the light field is also influenced
by a shift of the carrier wave with respect to the pulse peak
\cite{rmp}. Recently, the generation of intense, few-cycle laser pulses
with a stable carrier envelope phase $\varphi_{0}$ was demonstrated
\cite{b}. Let us consider the principle question for the design of
few-cycle pulse experiments: how does the pulse phase behave during
linear propagation? In order to answer this question, we can calculate
the evolution of few-cycle pulses in vacuum, which is most conveniently
described by a parabolic wave equation and by starting from a Gaussian
initial spatial pulse profile \cite{rmp}. Choosing the initial pulse
phase to be $\varphi_{0}$ at the beam waist, one reveals that the
carrier envelope phase in the far field $\varphi(\infty) = \varphi_{0}
- \pi/2$ undergoes a phase shift due to the Guoy phase shift $-\pi/2$
\cite{rmp}.  Note that the Guoy phase shift and all the other changes
experienced by the pulse during propagation do not depend on the
initial phase $\varphi_{0}$.

For an attosecond X-ray source it is of great interest to maximize the
central peak energy offset, which depends sensitively on the absolute
phase of the seed laser pulse $\varphi_{0}$. We start with an
illustration of the few-cycle-driven energy modulation assuming that
the peak electric field appears at the peak of the envelope when the
laser pulse passes the undulator center (i.e. $\varphi_{0} = 0$ at the
Gaussian beam waist). The interaction with the laser light in the
undulator then produces a time dependent electron energy modulation as
shown in Fig.~\ref{fig:emod40}.  For the laser (FWHM) pulse duration of
5 fs at a laser pulse energy 2-4 mJ, we expect a central peak energy
modulation 30-40 MeV.

\subsection{Monochromator}

The width of the spectral distribution of the SASE radiation will be
determined by the frequency chirp, provided the frequency chirp is
larger than FEL bandwidth. A monochromator may be used to select the
pulses of short duration, due to correlation between frequency and
longitudinal position in the radiation pulse. For 12 keV
photons, we consider Bragg diffraction in crystals as a method of
bandwidth selection. Special attention is called to the fact that the
relative spectral width for the given Bragg reflection is independent
of the wavelength or glancing angle of X-rays and is given merely by
properties of the crystal and the reflecting atomic planes.  In
particular, it implies that the choice of a crystal and reflecting
atomic planes determines the spectral resolution. For example, one can
consider Si(111) crystals, which have a FWHM bandwidth of
$\Delta\lambda/\lambda = 1.3\times 10^{-4}$, or Ge(111)crystals, which
have a FWHM bandwidth of $\Delta\lambda/\lambda = 3.4\times 10^{-4}$.
Monochromators at synchrotron beam lines are most commonly fabricated
from silicon. The reason is that the semiconductor industry has created
a huge demand for defect-free , perfect single crystals. Silicon is by
no means the only choice, and in recent years diamond has become a
popular alternative, due to the fact it has the highest thermal
conductivity and low absorption.

\begin{figure}[b]
\begin{center}

\vspace*{-5mm}

\epsfig{file=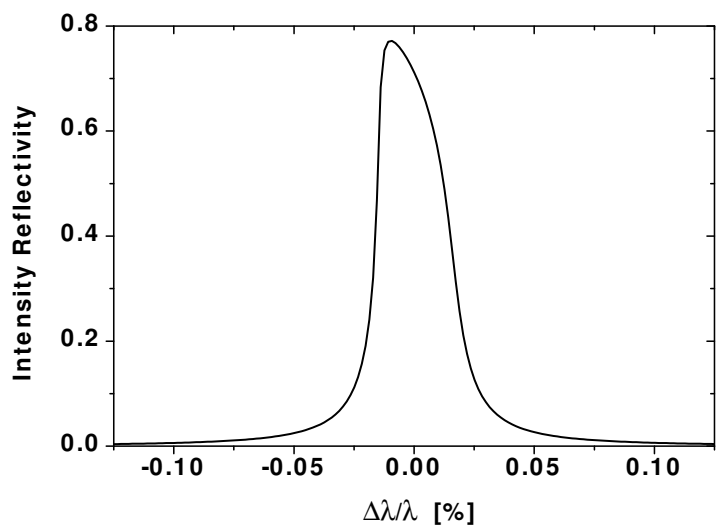,width=0.8\textwidth}

\vspace*{-5mm}

\end{center}
\caption{ Reflectivity curve for a
thick absorbing crystal in the Bragg case.  Germanium, 111, 0.1 nm }
\label{fig:ge1}
\end{figure}

An attosecond X-ray source requires a relatively broadband
monochromator.  The larger the monochromator bandwidth is, the shorter
the minimal pulse duration than can be extracted. We are discussing
here only Germanium single crystals which have the largest relative
bandwidth. Although Ge is not as perfect as silicon or diamond,
sufficiently large perfect Ge crystals are available today. For 12 keV
photons Bragg peaks of Ge crystals have reflectivities of approximately
75\%. Figure \ref{fig:ge1} gives an example of a reflectivity curve for
a thick absorbing crystal. The drawing of Fig. \ref{fig:ge1} shows
several interesting features. The shape is asymmetric and is due to
absorption effect. The tails of the reflectivity curve decrease as
$(\Delta\lambda/\lambda)^{-2}$. It should be pointed out that the tail
of reflectivity curve plays important role in the operation of the
attosecond X-ray source, and this characteristic of spectral window and
attosecond pulse contrast are ultimately connected. Good crystal
quality is required for high resolving power. Similarly, a good
resolving power requires a collimated beam.  In fact, the angular
spread of an insufficiently collimated beam negatively affects the
wavelength resolution just like poor crystal quality. The Ge
monochromator angular acceptance is of order 50 $\mu$rad for a
wavelength 0.1 nm, and is well matched to the natural opening angle,
(1~$\mu$rad), of an XFEL source. Therefore, a crystal monochromator
works better with XFEL radiation than with conventional synchrotron
source.

Besides the crystal quality, other factors must be considered in
selecting a scheme for a monochromator. The monochromator crystal must
be thermally stable and capable of being exposed to XFEL output
radiation with limited radiation damage. We have chosen the one Ge
(111) crystal scheme of the X-ray monochromator with silicon
premonochromator, which withdraws the major heat load from the actual
short pulse selection Ge monochromator. We consider Laue diffraction in
Si crystals as a method of bandwidth selection in premonochromator.  In
the pre-monochromator part one can use ten Si crystal plates of $15\mu
$m thickness and the reflection Si(111). Given the crystal plate is
perfect, it reflects 90\% of the incident X-rays within a band of
$\Delta\lambda/\lambda \simeq 10^{-4}$. One can align the Si plates so
that the main peak of the spectrum is blocked. The radiation power
which reaches the broadband (Ge) monochromator crystal is 10\% of the
initial value. Only 30\% of the offset frequency radiation is absorbed
in ($15\times 10 = 150 \mu$m thickness) premonochromator totally, and
the rest passes through. Another advantage of the premonochromator is
the possibility to increase the contrast of output attosecond X-ray
pulses.

\subsection{Output characteristics of attosecond FEL}

A complete description of the X-ray FEL can be performed only with
three-dimensional, time-dependent numerical simulation code. Since
amplification process starts from shot noise, properties of a
single-spike selection should be described in statistical term.  The
statistics of concern are defined over an ensemble of radiation pulses.

\begin{figure}[tb]

\begin{center}

\vspace*{-5mm}

\epsfig{file=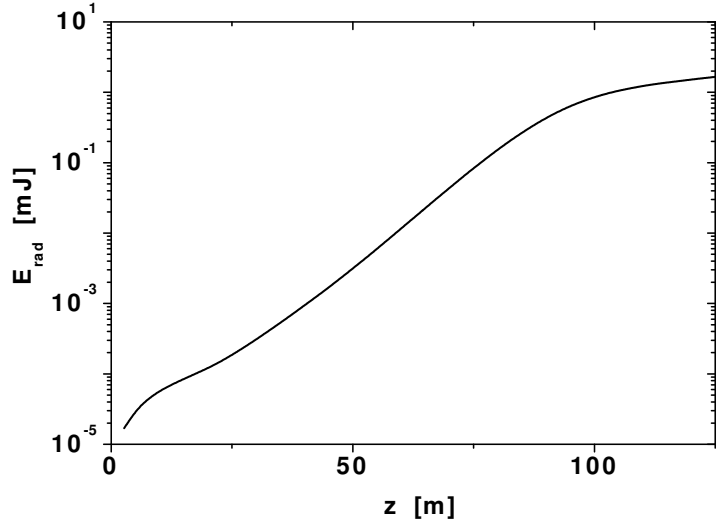,width=0.8\textwidth}

\vspace*{-5mm}

\end{center}

\caption{
Average energy in the radiation pulse versus undulator length
}
\label{fig:pz40}
\end{figure}

\begin{figure}[tb]

\vspace*{-5mm}

\begin{center}
\epsfig{file=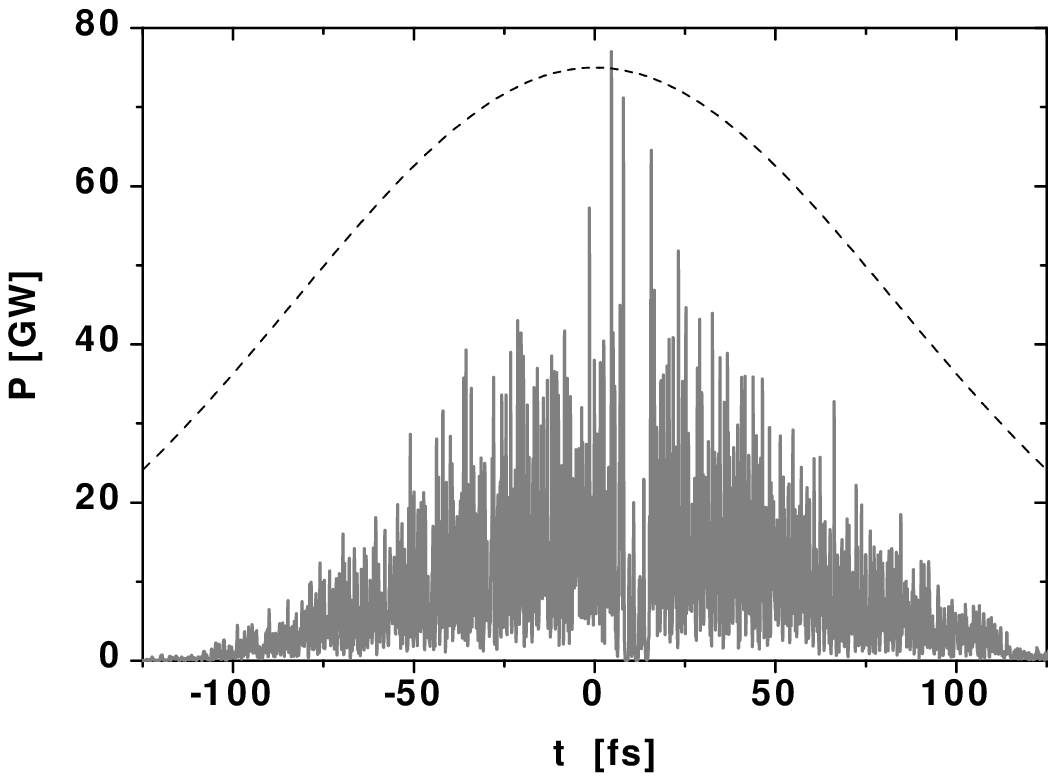,width=0.8\textwidth}

\vspace*{-10mm}

\epsfig{file=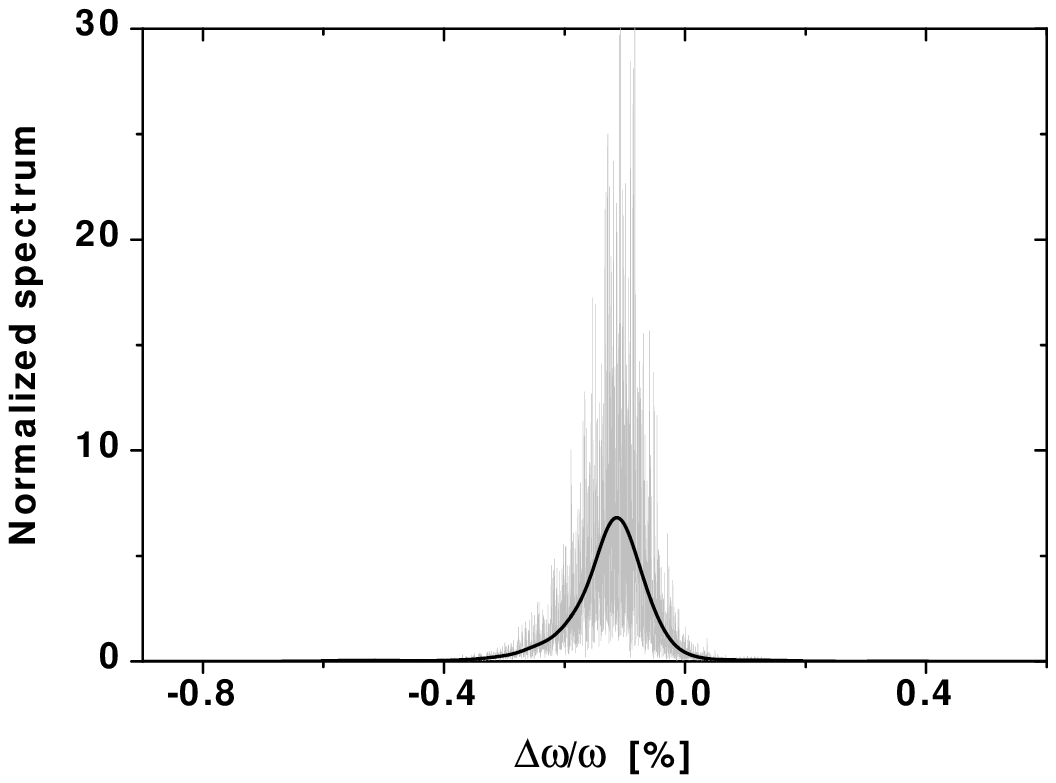,width=0.8\textwidth}
\end{center}

\caption{
Temporal (upper plot) and spectral (lower plot) structure of the
radiation pulse. Solid line at the lower plot shows averaged spectrum.
Undulator length is 120~m
}
\label{fig:ps0192050}
\end{figure}

\begin{figure}[tb]

\vspace*{-5mm}

\begin{center}
\epsfig{file=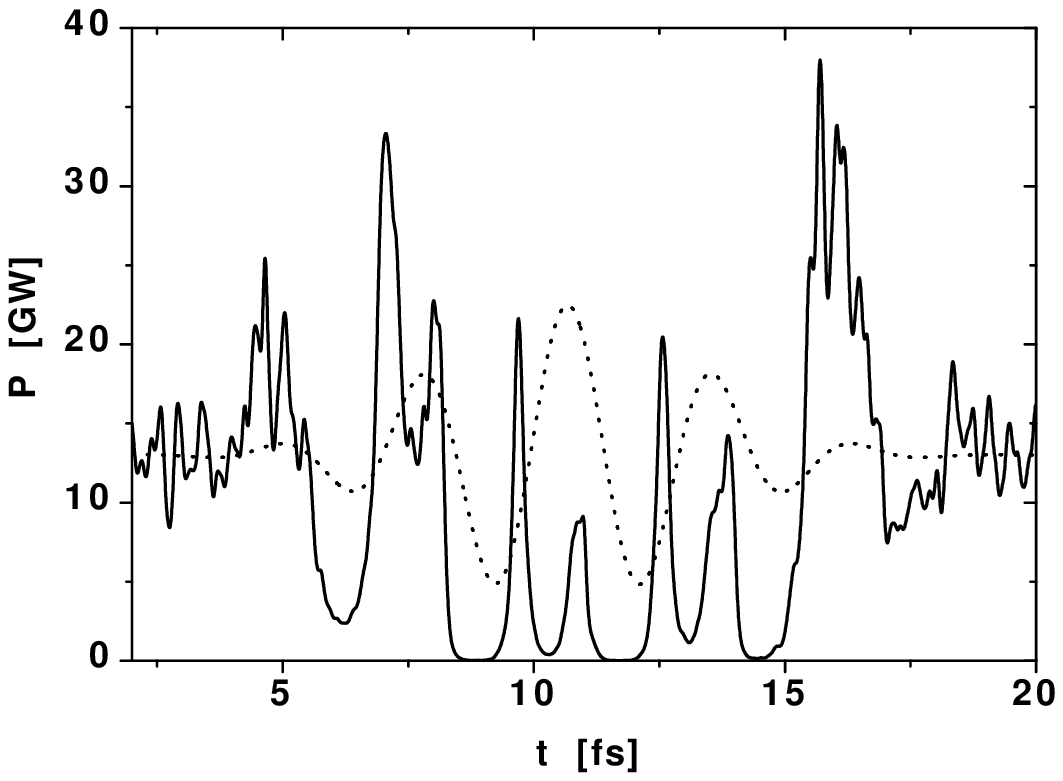,width=0.8\textwidth}

\vspace*{-10mm}

\epsfig{file=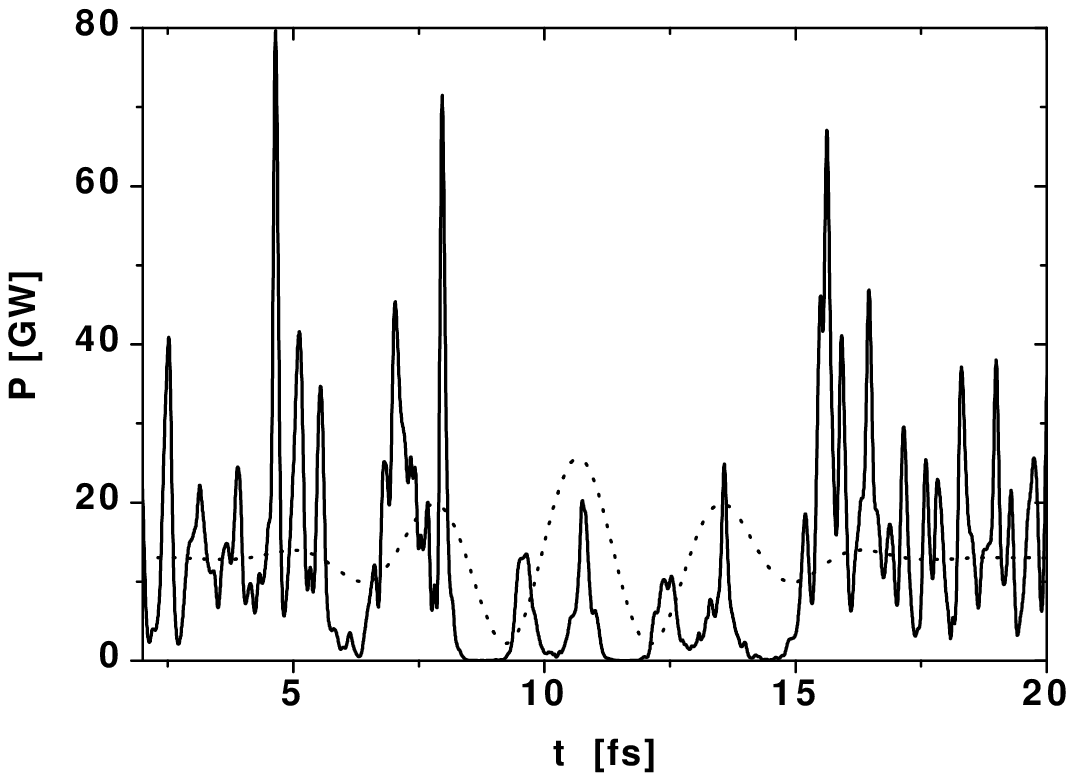,width=0.8\textwidth}
\end{center}

\caption{
Averaged (upper plot) and typical single-shot (lower plot)
temporal structure of the central part of the radiation pulse.
Undulator length is 120~m.
Dotted line shows energy modulation of the electron bunch
(see Fig.~\ref{fig:emod40})
}
\label{fig:pav4050}
\end{figure}

\begin{figure}[tb]

\vspace*{-5mm}

\begin{center}
\epsfig{file=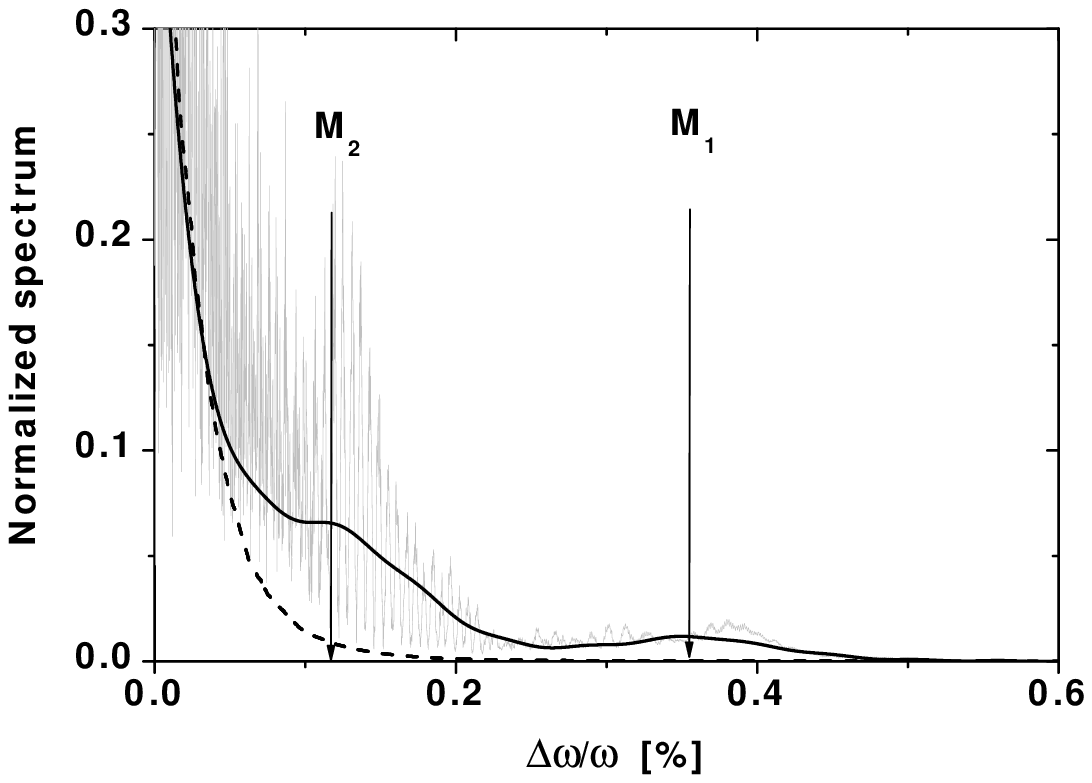,width=0.8\textwidth}

\vspace*{-10mm}

\epsfig{file=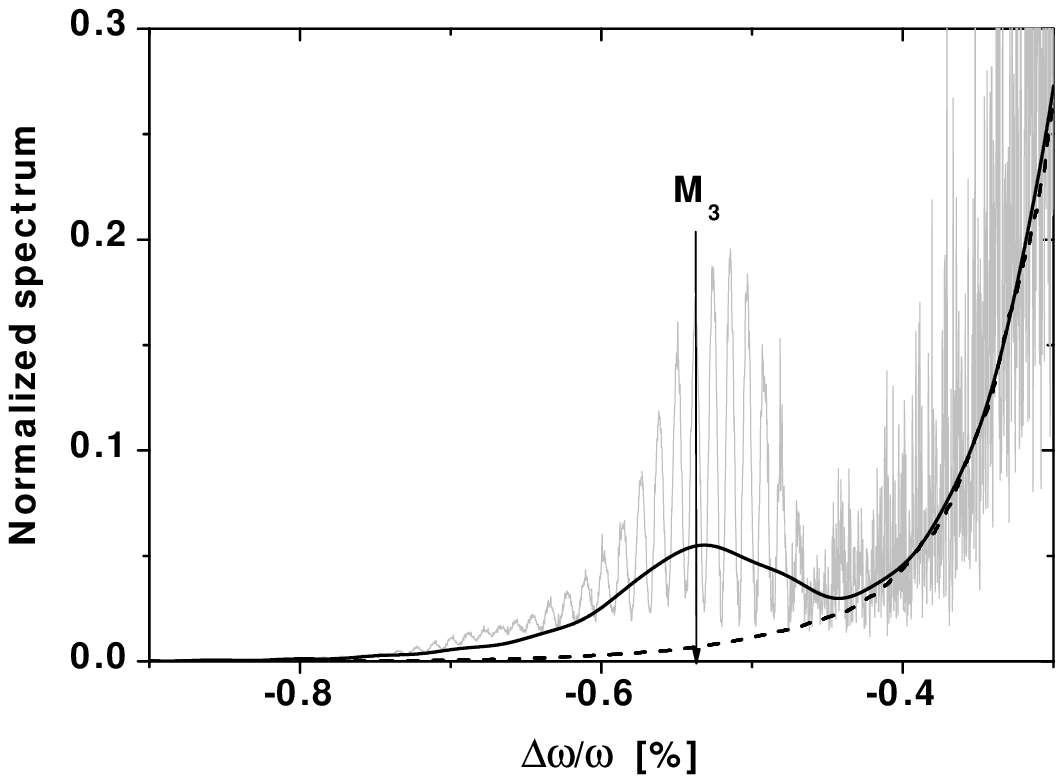,width=0.8\textwidth}
\end{center}

\caption{
Spectrum of the radiation pulse produced by modulated electron
bunch (gray line).
Undulator length is 120~m.
Plots show enlarged tails of
complete spectrum presented in Fig.~\ref{fig:ps0192050}.
Solid line is averaged spectrum.
Dashed line is averaged spectrum of nonmodulated electron beam.
Mark $M_1$ shows tuning of monochromator for single pulse selection
(see Fig.~\ref{fig:q4050}). Marks $M_2$ and $M_3$ show tuning of
the monochromator for selection of two pulse sequence
(see Figs.~\ref{fig:r4050} and \ref{fig:l4050}).
}
\label{fig:s0192050a}
\end{figure}

\begin{figure}[tb]

\vspace*{-5mm}

\begin{center}
\epsfig{file=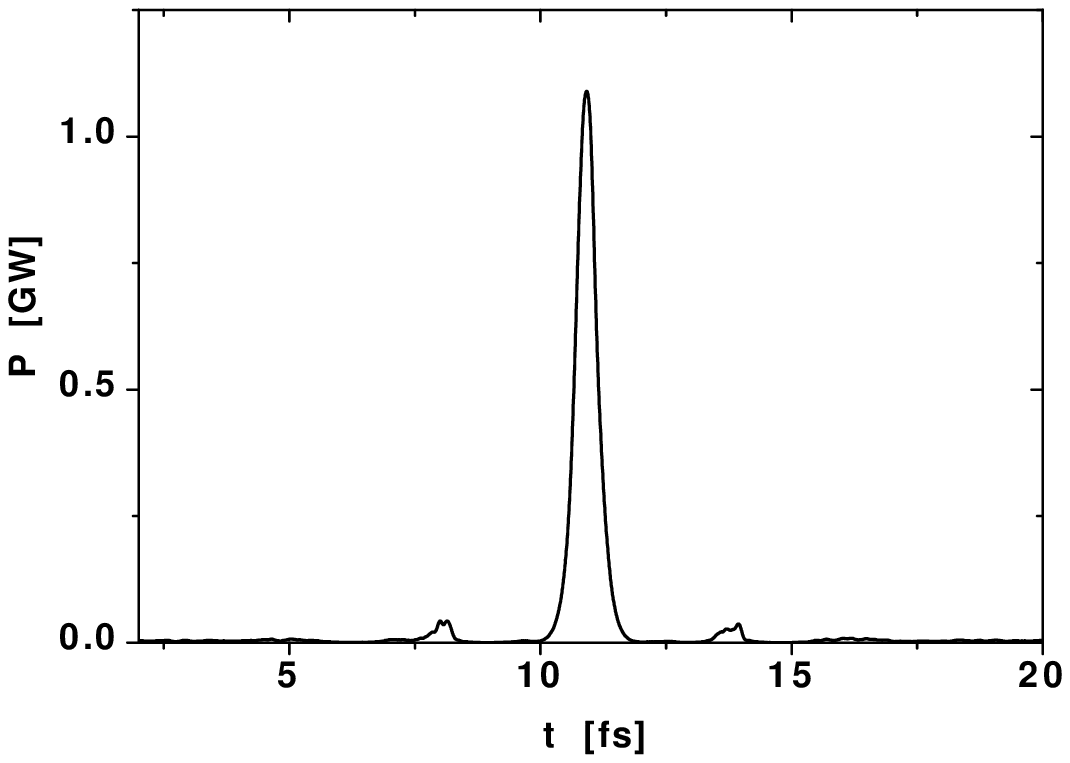,width=0.8\textwidth}

\vspace*{-10mm}

\epsfig{file=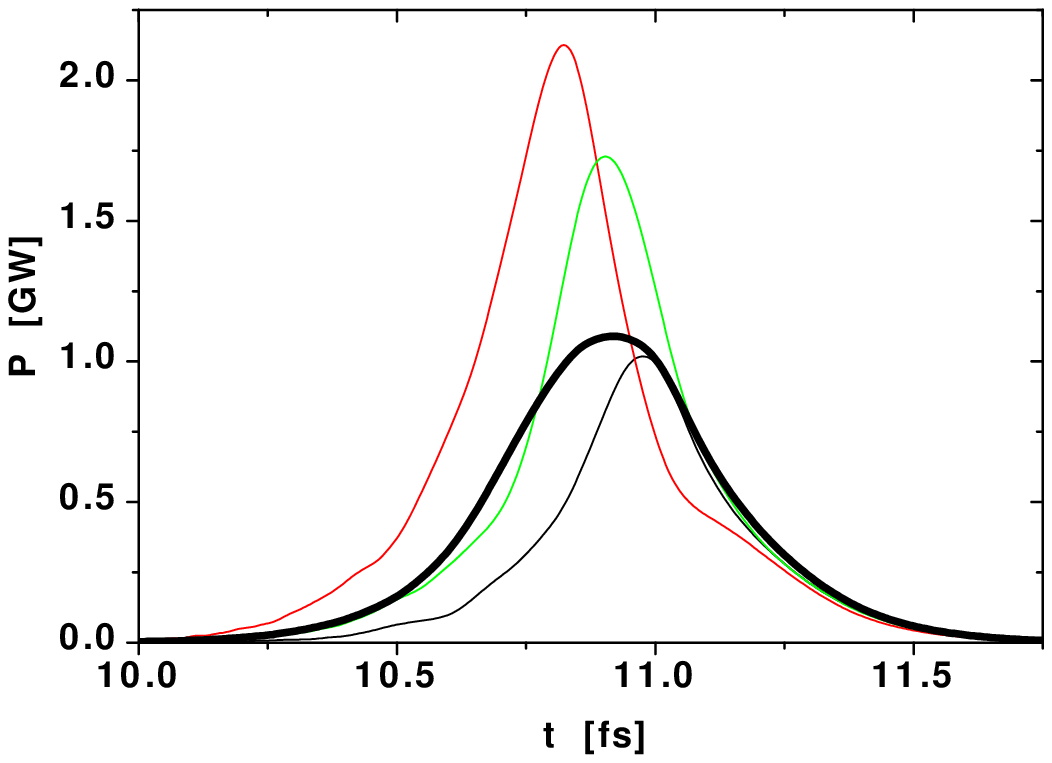,width=0.8\textwidth}
\end{center}

\caption{
Temporal structure of the radiation pulse behind
monochromator tuned to single spike selection
(mark $M_1$ in Fig.~\ref{fig:s0192050a}).
Upper plot shows average over many pulses, and lower plot
shows details of single pulses.
Bold curve is average over many pulses
}
\label{fig:q4050}
\end{figure}

\begin{figure}[tb]

\vspace*{-5mm}

\begin{center}
\epsfig{file=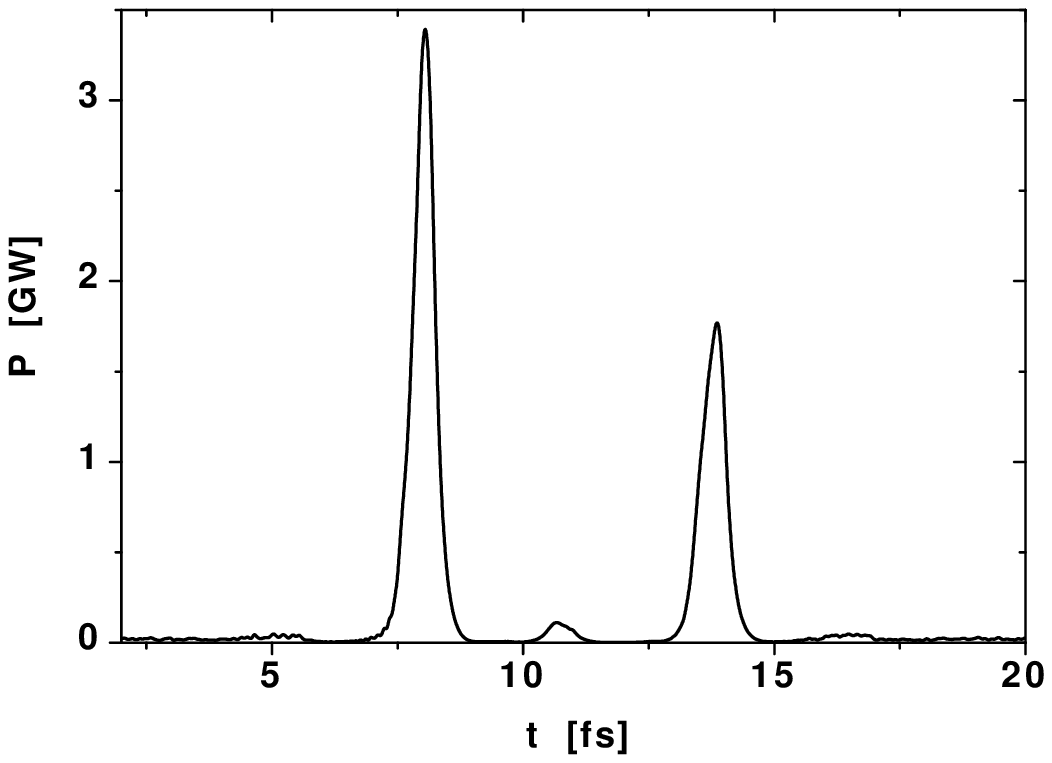,width=0.8\textwidth}

\vspace*{-10mm}

\epsfig{file=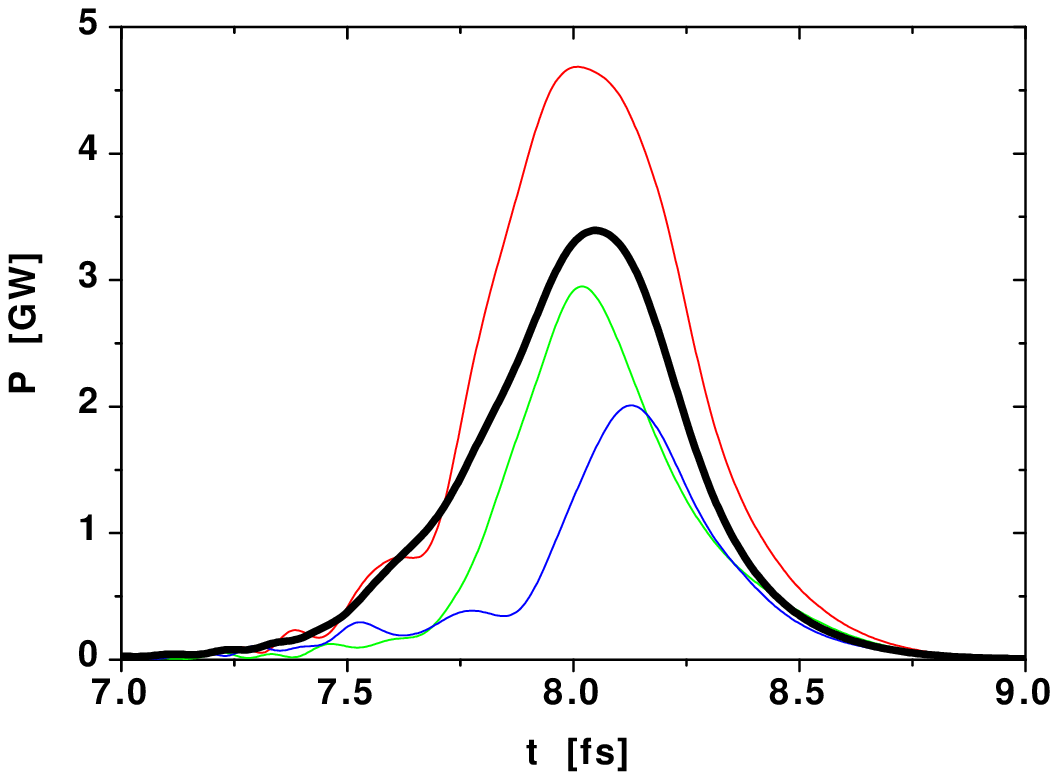,width=0.8\textwidth}
\end{center}

\caption{
Temporal structure of the radiation pulse behind
monochromator tuned to selection of two pulse sequence
(mark $M_2$ in Fig.~\ref{fig:s0192050a}).
Pulse separation is two laser oscillation periods.
Upper plot shows average over many pulses, and lower plot
shows details of single pulses.
Bold curve is average over many pulses
}
\label{fig:r4050}
\end{figure}

\begin{figure}[tb]

\vspace*{-5mm}

\begin{center}
\epsfig{file=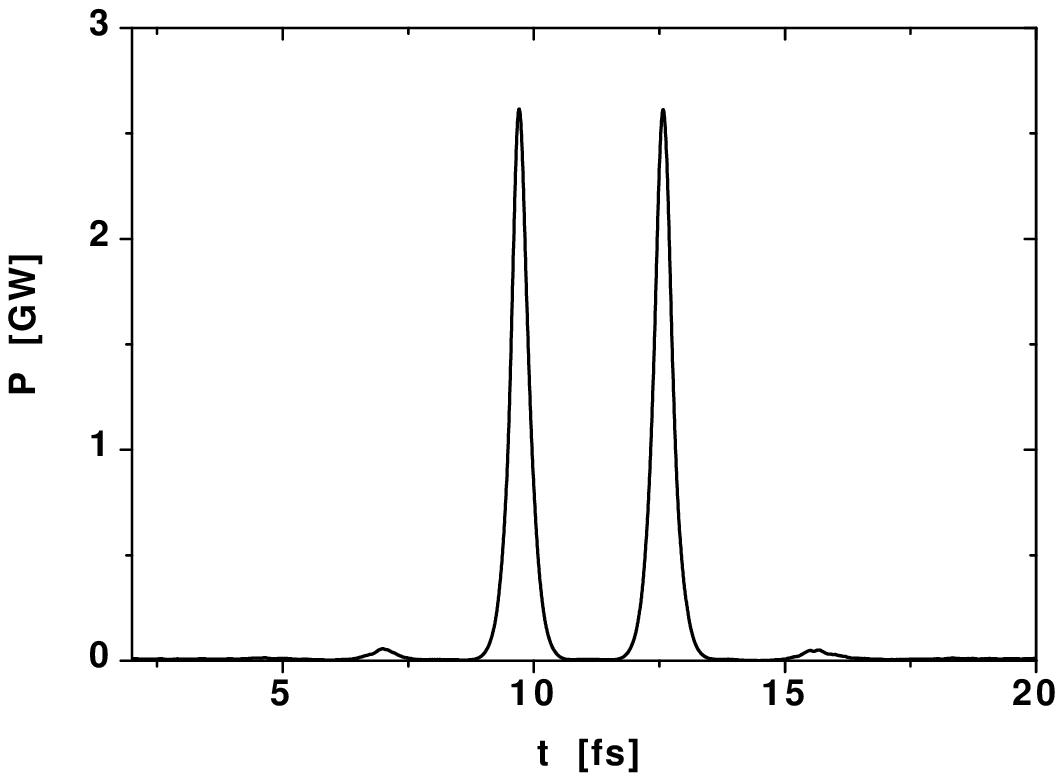,width=0.8\textwidth}

\vspace*{-10mm}

\epsfig{file=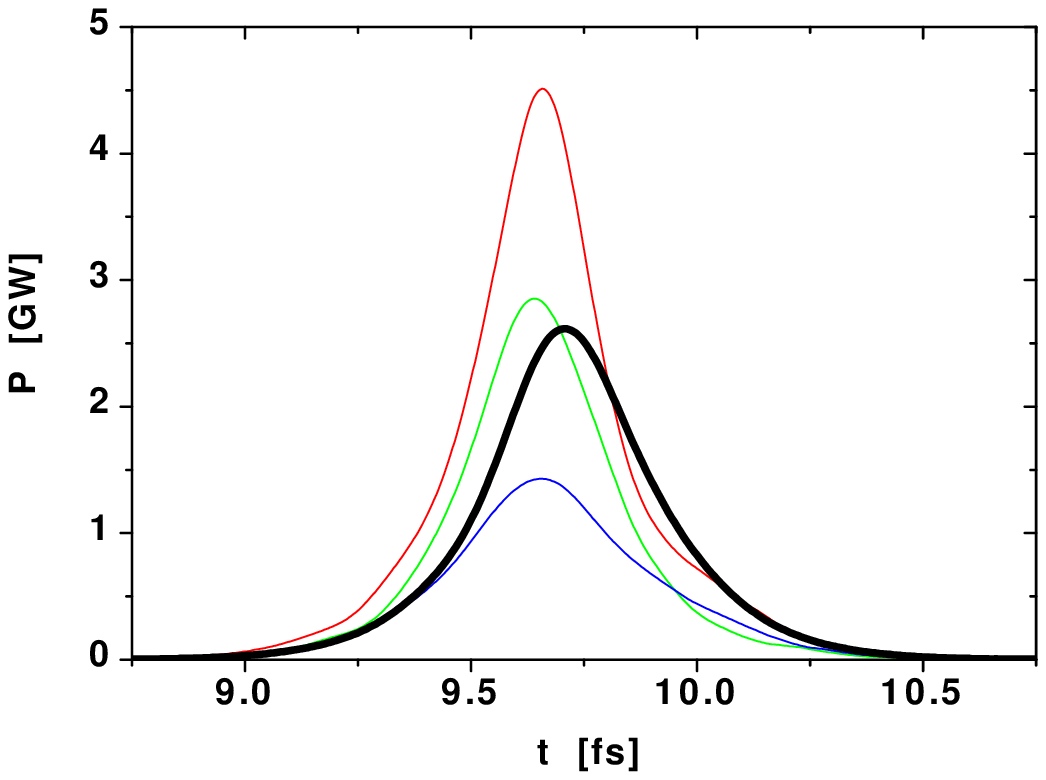,width=0.8\textwidth}
\end{center}

\caption{
Temporal structure of the radiation pulse behind
monochromator tuned to selection of two pulse sequence
(mark $M_3$ in Fig.~\ref{fig:s0192050a}).
Pulse separation is one laser oscillation period
Upper plot shows average over many pulses, and lower plot
shows details of single pulses.
Bold curve is average over many pulses
}
\label{fig:l4050}
\end{figure}

In the present scheme an electron beam with slice modulation of the
energy passes through the undulator and produces SASE radiation. Since
only a small fraction of the bunch is modulated (10~fs versus 200~fs of
FWHM electron pulse duration, see Fig.\ref{fig:emod40}), the total
energy in the radiation pulse remains approximately the same as in the
case of nonmodulated electron beam (see Fig.~\ref{fig:pz40}), and
saturation is achieved at an undulator length of about 120~m.
Figure~\ref{fig:ps0192050} shows typical temporal and spectral
structure of the radiation pulse at saturation. In the present
numerical example the central part of the electron bunch was subjected
to the slice energy modulation, and one can notice its clear signature
in the temporal structure. Figure~\ref{fig:pav4050} shows an enlarged
view of the central part of the radiation pulse. The dotted lines in
this figure show the initial energy modulation of the electron beam.
The temporal structure of the radiation pulse has a clear physical
explanation. The FEL process is strongly suppressed in the regions of
the electron bunch with large energy chirp, and only regions of the
electron bunch with nearly zero energy chirp produce radiation. From a
physical point of view each part of the bunch near the local extremum
of the energy modulation can be considered as an isolated electron
bunch of short duration. At the chosen parameters of the system its
duration is about 300~attosecond which is about of coherence time.
Thus, it is not surprising that only a single radiation spike is
produced by each area near the local extremum. An asymmetry of averaged
power (at symmetric energy modulation) is due to the nonsymmetry of the
FEL process with respect to the sign and the value of the energy chirp.
In particular, spikes related to the negative of energy offset have a
higher amplitude. This is a typical nonlinear effect allowing to
prolong interaction of the radiation pulse with the electron pulse at
the "correct" sign of the energy chirp. The temporal structure of the
radiation pulse nearly repeats the temporal structure of the energy
modulation. Slight deviations from the periodic structure are also due
to the nonsymmetry of the FEL process with respect to the sign of the
energy chirp. In particular, spikes related to the negative energy
offset slip more forward than those related to the positive energy
offset.

Let us turn back to the main subject of our study, i.e. to the
production of attosecond pulses. The lower plot in
Fig.~\ref{fig:ps0192050} shows the total radiation spectrum of the
radiation pulse. At this scale a signature of the slice energy
modulation can hardly be seen. In Fig:~\ref{fig:s0192050a} we present
the tails of the spectrum at an enlarged scale. Each of three clearly
visible bumps in the averaged spectrum corresponds to a local extremum
of the energy offset shown in Fig.~\ref{fig:emod40}. The bump marked as
$M_1$ corresponds to the central peak energy offset. The bump $M_2$
corresponds to the neighboring two positive energy offsets. The bump
$M_3$ comes from the areas of the electron bunch with negative energy
offset. The single-shot spectrum (shown as grey line) exhibits an
oscillatory behavior near bumps $M_2$ and $M_3$. That is due to an
interference of two radiation wavepackets with close frequencies coming
from different parts of the electron bunch. Other maxima can be hardly
distinguished, since they are located within the bandwidth of the main
spectrum.

Figures~\ref{fig:xas2}, \ref{fig:xas3}, and \ref{fig:s0192050a} give a
clear idea about separation of the attosecond radiation pulses.
Positioning of the monochromator to different maxima of the spectrum
allows us to select single pulse, or a two pulse sequence of attosecond
duration. The calculation involves the following steps. The FEL
simulation code produces 3-D arrays for the radiation field in the near
zone. This field is recalculated into the far field zone, and is
subjected to the Fourier transform. The latter result is convoluted
with the reflectivity function of Ge(111) monochromator (see
Fig.~\ref{fig:ge1}), and is subjected to inverse Fourier transform
giving temporal structure of the radiation pulse behind the
monochromator.

By selecting the frequency offset of the monochromator to the position
marked as $M_1$ in Fig.~\ref{fig:s0192050a}, we select single pulses.
Their properties are illustrated with Fig.~\ref{fig:q4050}. An analysis
of single pulses shows that their pulse duration is about 300~as, the
average power has GW-level, and the radiation pulse energy is about a
$\mu $J. The larger width of the averaged curve is partially due to
shot-to-shot fluctuations of the position of the radiation pulse (a
fraction of coherence time). Note that shot-to-shot fluctuations of the
radiation energy after monochromator are suppressed significantly due
to ultrashort duration of the lasing fraction of the electron bunch
\cite{fel2002-short-bunch}.  An advantage of single-pulse selection is
the small background from the main radiation pulse due to a large
offset from the resonant frequency.

By positioning of the monochromator central frequency to the spectrum
bumps $M_2$ or $M_3$ one can select a two pulse sequence as illustrated
in Figs.~\ref{fig:r4050} and \ref{fig:l4050}. Two pulses are separated
by two or one oscillation period of optical laser depending on the
choice of the monochromator tuning. Note that due to the statistical
nature of the SASE process the time jitter between two pulses is about
200~as, a fraction of the coherence time. One should not wonder that
pulse amplitudes differ visibly for the case of pulse separation by one
laser oscillation period (see Fig.~\ref{fig:r4050}). As mentioned
above, this is a typical nonlinear effect related to the sensitivity of
the FEL process to the sign and the value of the energy chirp. Although
the energy modulation amplitude is the same in both maxima, the shape
of the energy chirp is asymmetric.

\section{Discussion}

Successful operation of the attosecond XFEL requires the fulfillment of
several requirements. The requirement that the SASE FEL bandwidth is
much less than the separation of the few-cycle-driven frequency offset
is of critical importance for the performance of the attosecond XFEL.
In this case a crystal monochromator can be used to distinguish the
attosecond pulses from the intense SASE pulses. Obviously, this
requirement is easier to achieve for high power optical laser systems.
For 800 nm laser radiation and for 0.1 nm output radiation, for
example, the peak power of few-cycle laser pulse must be larger than
500-700 GW. This condition can be satisfied by a terawatt-scale sub-10
fs Ti:sapphire laser system which seems feasible.

Our scheme of attosecond X-ray source is based on the assumption that
the beam density modulation does not appreciably change as the beam
propagates through the energy modulator undulator. When the resonance
condition takes place, the electrons with different arrival phases
acquire different values of the energy increments (positive or
negative), which result in the modulation of the longitudinal velocity
of the electrons with the laser frequency.  Since the velocity
modulation is transformed into a density modulation of the electron
beam when, passing the undulator, an additional wakefield exists
because of a variation in amplitude density modulation. It is
interesting to estimate the amount of bunching produced during the 800
nm undulator pass. An undulator is a sequence of bending magnets where
particles with different energies have different path length, $\Delta z
= R_{56}\delta E/E$. The net compaction factor of the undulator is given
by $R_{56} = 2\lambda_{0}N_{\mathrm{w}}$, where $\lambda_{0} = 800$~nm
is the resonance wavelength and $N_{\mathrm{w}} = 2$ is the number of
undulator periods. An induced correlated energy spread at the exit of
(800 nm) undulator is about 0.3\%. Therefore, a rough estimate for the
induced bunching is $\delta a \simeq (\pi R_{56}/\lambda_{0})(\delta
E/E) \simeq 3 \times 10^{-2}$. Since this value is much less than
unity, we can conclude that density modulation in the 800 nm undulator
due to few-cycle-driven energy spread should not be a serious
limitation in our case.

The next problem is that of synchronization.  Frequency chirp in the
XFEL is seeded by positioning a fs optical pulse on the electron bunch.
Even when femtosecond pulses from laser system are synchronized to the
photoinjector master clock with phase-locking technique, the
synchronization of the optical seed laser with the electron pulses to
an accuracy of 100 fs is not yet achievable. A more serious problem is
the timing jitter of electron and seed laser pulses. The jitter of
electron pulses originates in the photoinjector laser system (laser
pulse jitter) and in the magnetic bunch compressors (from electron
bunch energy jitter). Due to this uncertainty, not every fs optical
pulse will produce an attosecond X-ray pulse.
Random production of attosecond X-ray pulses needs to be
controlled. A basic question at this point is how attosecond X-ray
pulses will be identified. Separation of attosecond pulse frequency
from the central frequency can be used to distinguish the 300 as pulses
from the intense 100 fs pulses. Appearing of X-ray pulses at the
frequency offset will indicate that the seed optical pulse is
overlapped with the central part of the electron bunch.

Analysis of parameters of an attosecond X-ray source shows that its
repetition rate is clearly limited by the value of repetition rate
achievable with terawatt-scale sub-10 fs Ti:sapphire laser system
having 1-kHz repetition rate \cite{b}. The single-pass scheme
considered here is the simplest one. The laser beam, which is
essentially unaltered in the electron beam modulation process, is then
disposed of.  This is not optimum for a couple of reasons. The idea is
roughly the following. The attosecond pulse repetition rate can be
significantly increased if the laser pulse can be reused, because the
laser pulse suffers little loss in energy after each interaction with
electron beam. The solution of this problem is a multipass approach
(based on reflective optical elements) in which a laser pulse is made
to pass through the modulator undulator a finite number of times before
being thrown away.  In this way, the attosecond pulse repetition rate
is increased by increasing the number of laser pulses used.

\section{Conclusion}

Operation of the proposed scheme was illustrated for the parameters of
the European XFEL. Although the present work is concerned primarily for
use in the wavelength range around 0.1 nm, its applicability is not
restricted to this range, for example 0.15 nm LCLS facility is a
suitable candidate for application of attosecond techniques described
here. It is important that proposed attosecond scheme is based on the
nominal XFEL parameters, and operates in a "parasitic" mode not
interfering with the main mode of the XFEL operation. It can be
realized with minimum additional efforts. The machine design should
foresee the space for installation of modulator undulator and a
viewport for input optical system. Many of the components of the
required laser system can be achieved with technology which is
currently being developed for applications other than the attosecond
X-ray source. As a result,  a laser system could be developed over the
next few years and can meet the XFEL requirements well in advance of
XFEL construction schedule.

\section*{Acknowledgments}

We thank G.~Gr\"ubel, D.~Novikov, E.~Weckert  for many useful
discussions.  We thank R.~Brinkmann, J.R.~Schneider, A.~Schwarz, and
D.~Trines for interest in this work.

\end{document}